\documentclass{webofc}
\usepackage[varg]{txfonts}   % Web of Conferences font
\usepackage{hyperref}
\usepackage{url}
\hypersetup{colorlinks=true,citecolor=blue,urlcolor=blue,linkcolor=blue}
\begin{document}
\title{Compositeness of near-threshold exotic hadrons in systems with Coulomb and short-range interactions}

\author{\firstname{Tomona} \lastname{Kinugawa}\inst{1}\fnsep\thanks{\email{tomona.kinugawa@riken.jp}} \and
        \firstname{Tetsuo} \lastname{Hyodo}\inst{2}\fnsep\thanks{\email{hyodo@rcnp.osaka-u.ac.jp}}
}

\institute{Nishina Center for Accelerator-Based Science, RIKEN, Wako 351-0198, Japan 
\and
           Research Center for Nuclear Physics, The University of Osaka, Ibaraki, Osaka 567-0047, Japan 
          }

\abstract{
We study the universal properties of near-threshold $s$-wave eigenstates in systems with Coulomb plus short-range interactions by analyzing their compositeness. Using a nonrelativistic effective field theory, we derive a relation between the compositeness of near-threshold states and the Coulomb scattering length, the Coulomb effective range, and the Bohr radius. We find that the long-range Coulomb interaction modifies the threshold behavior expected from the low-energy universality of purely short-range systems. When the Coulomb interaction is relatively weak compared with the short-range interaction, a remnant of the low-energy universality associated with the short-range interaction survives, leading to composite-dominant near-threshold eigenstates. In contrast, in the strong-Coulomb regime, this remnant disappears.
}
\maketitle
\section{Introduction}
\label{intro}

% exotic hadrons
In recent years, many candidates for exotic hadrons have been observed near two-hadron scattering thresholds~\cite{ParticleDataGroup:2024cfk}. Famous examples are the $X(3872)$ and $T_{cc}$, which are near-threshold states with very small binding energies. The internal structure of exotic hadrons has been intensively studied, and various configurations, such as hadronic molecules and compact multiquark states, have been considered~\cite{Hosaka:2025gcl}.

%Hosaka:2016pey,Brambilla:2019esw,

% compositeness
The internal structure of near-threshold states can be quantitatively characterized by the compositeness $X$~\cite{Weinberg:1965zz,Hyodo:2013nka,Oller:2017alp,vanKolck:2022lqz,Kinugawa:2024crb}. For a stable bound state, the compositeness corresponds to the probability of finding the hadronic molecular component in the wave function of the state. The remaining probability is called the elementarity $Z$. In general, the compositeness characterizes the clustering structure of the state. For instance, a hadronic molecule can be induced by the clustering of quarks. Owing to this feature, the compositeness has been widely used to investigate the internal structure of exotic hadrons~\cite{Kinugawa:2024crb}, while the same concept has also been discussed in nuclear and atomic systems~\cite{Kinugawa:2022fzn}.

% short-range interaction
For an $s$-wave system with only a short-range interaction, the structure of a near-threshold state is constrained by low-energy universality~\cite{Braaten:2004rn,Naidon:2016dpf}. In particular, in the limit where the binding energy goes to zero, a bound state becomes purely composite with $X \to 1$~\cite{Hyodo:2014bda}. As a consequence of this fact, shallow bound states are usually composite dominant~\cite{Kinugawa:2023fbf}. On the other hand, near-threshold narrow resonances with a small excitation energy tend to be non-composite dominant~\cite{Kinugawa:2024kwb}. This universal behavior provides a theoretical basis for the empirical expectation that molecular and cluster states appear near scattering thresholds.

% Coulomb + short-range interaction
However, the constituents of cluster states can also interact through the long-range Coulomb interaction in addition to short-range interactions, such as the nuclear force. Although the electromagnetic interaction is weaker than the strong interaction, its effect can become important for shallow near-threshold states. For example, the ${}^{8}$Be nucleus appears as a near-threshold resonance in the $\alpha\alpha$ system, while it becomes a bound state when the Coulomb repulsion is switched off~\cite{Higa:2008dn}. In the case of an attractive Coulomb interaction, the short-range interaction in the $\Xi^{-}\alpha$ system is not sufficiently attractive to form a bound state by itself, and the Coulomb attraction assists the formation of a bound state~\cite{Hiyama:2022jqh}. These examples indicate that the Coulomb interaction can qualitatively modify the threshold behavior and the nature of the eigenstates.

% this work
In this work, we investigate the compositeness of near-threshold $s$-wave eigenstates in systems with Coulomb plus short-range interactions. Using a nonrelativistic effective field theory, we study the pole trajectories and the compositeness, and clarify how the low-energy universality is modified by the Coulomb interaction.  

%%%%%%%%%%%%%%%%
\section{Formulation}

We consider an $s$-wave two-body system with the Coulomb plus short-range interactions using a nonrelativistic effective field theory~\cite{Kong:1998sx,Kong:1999sf,Ando:2007fh,Higa:2008dn}. The full scattering amplitude consists of the pure-Coulomb contribution and the Coulomb-distorted short-range contribution. The latter determines the pole condition for the eigenstates and is written as
\begin{align}
f_{CS}(k) &= C_{\eta}^{2}e^{2i\sigma_{0}}\left[-\frac{1}{a^{C}_{s}} + \frac{r^{C}_{e}}{2}k^{2} - 2k_{C}H(\eta) \right]^{-1}.
\label{eq:fCS-renorm}
\end{align}
where $a_{s}^{C}$ and $r_{e}^{C}$ are the Coulomb scattering length
and the Coulomb effective range, respectively. Here,
$\eta=k_{C}/k$ with $k_{C}=\mu\alpha Z_{1}Z_{2}$.
The Bohr radius is defined as
\begin{equation}
a_{B}=\frac{1}{|k_{C}|}.
\end{equation}
To characterize the internal structure of the eigenstates, we introduce the compositeness $X$. In the present scheme, the compositeness is expressed as
\begin{align}
X &= \left[1 - \frac{r_{e}^{C}}{R^{C}}\right]^{-1}, 
\label{eq:wbr}
\end{align}
where $R^{C}$ is the length scale associated with the bound-state wave function in the presence of the Coulomb interaction. The detailed formulation and the explicit expression for $R^{C}$ are given in Ref.~\cite{Kinugawa:2026fob}.

%%%%%%%%%%%%%%%%
\section{Universal properties of near-threshold eigenstates}

% pole trajecotry
In Ref.~\cite{Kinugawa:2026fob}, we first investigated the pole trajectories of near-threshold eigenstates in systems with Coulomb plus short-range interactions by fixing $r_{e}^{C}$ and $a_{B}$ and varying $a_{s}^{C}$. For a repulsive Coulomb interaction, the bound-state pole is continuously connected to a resonance pole across the threshold, in contrast to a purely short-range system, where the bound-state pole turns into a virtual-state pole. For an attractive Coulomb interaction, the bound-state pole originating from the short-range interaction approaches the pure Coulomb ground-state level and is not continuously connected to the resonance trajectory. This behavior is qualitatively different from both the purely short-range and repulsive-Coulomb cases because infinitely many Coulomb bound-state poles accumulate near the threshold.

% scales
To understand the behavior of the compositeness along these pole trajectories, we next consider the characteristic scales of the system. The characteristic length scales associated with the Coulomb and short-range interactions are given by $a_{B}$ and $|r_{e}^{C}|$, respectively. The Bohr radius $a_{B}$ sets the length scale at which the Coulomb interaction becomes important, and therefore the Coulomb contribution cannot be neglected in the momentum region $k\lesssim 1/a_{B}$. On the other hand, in a purely short-range system, low-energy universality emerges in the region $k\ll 1/|r_{e}|$, where the effective-range term in the scattering amplitude is negligible, and the scattering amplitude is governed mainly by the scattering length. Therefore, a remnant of the short-range universality can appear if there exists an intermediate momentum region satisfying
$1/a_{B} \ll k \ll 1/|r_{e}^{C}|$. Thus, whether such a region exists is determined by the ratio $|r_{e}^{C}|/a_{B}$.

% weak Coulomb
In the weak-Coulomb regime, $|r_{e}^{C}|/a_{B}\ll1$, such an intermediate momentum region exists. In this region, the pole position is mainly governed by the scattering length, and a remnant of the low-energy universality of purely short-range systems survives. As a consequence, the compositeness of near-threshold bound states is enhanced toward unity, and the states become composite dominant.

% weak Coulomb + repulsive
Furthermore, for a repulsive Coulomb interaction, the continuity of the pole trajectory across the threshold indicates that the composite-dominant nature of the bound state in the weak-Coulomb regime persists into the resonance region. Since the compositeness of a resonance is in general complex, a prescription is required to interpret its internal structure. By applying the interpretation prescription introduced in Ref.~\cite{Kinugawa:2026fob}, we find that near-threshold resonances can also be composite dominant.

% weak Coulomb + attractive
For an attractive Coulomb interaction, the compositeness of the bound state approaches unity in the limit corresponding to the pure Coulomb ground state, regardless of the effective range. This is because the pure Coulomb ground state has no bare-state component and hence has $X=1$. However, the compositeness approaches unity more rapidly in the weak-Coulomb regime, $|r_e^C|/a_B\ll1$. This faster enhancement of the compositeness can be understood as a remnant of the short-range universality. The detailed pole trajectories and compositeness for both attractive and repulsive Coulomb interactions are discussed in Ref.~\cite{Kinugawa:2026fob}.

% strong Coulomb
On the other hand, when $|r_{e}^{C}|/a_{B}\gg1$, corresponding to the strong-Coulomb regime, the Coulomb-dominant region is reached before the short-range universal region can appear. In this case, the Coulomb interaction cannot be neglected even in the region where the effective-range contribution becomes small. Consequently, the enhancement of the compositeness associated with the short-range universality disappears, and near-threshold states are no longer necessarily composite dominant.

%%%%%%%%%%%%%%%%
\section{Summary}

We have investigated the compositeness of near-threshold $s$-wave eigenstates in systems with Coulomb plus short-range interactions. The long-range Coulomb interaction modifies both the pole structure and the internal structure of the eigenstates compared with systems governed only by the short-range interactions. The relative importance of the Coulomb interaction near the threshold is characterized by the ratio $|r_{e}^{C}|/a_{B}$. When $|r_{e}^{C}|/a_{B} \ll 1$, a remnant of the short-range universality survives, and near-threshold eigenstates tend to be composite dominant. In particular, for the repulsive Coulomb interaction, a bound state is continuously connected to a resonance across the threshold, and the near-threshold resonance can also be composite dominant. On the other hand, when $|r_{e}^{C}|/a_{B} \gg 1$, the enhancement of the compositeness near the threshold disappears, and near-threshold states are not necessarily composite dominant and can be non-composite dominant.

% BibTeX or Biber users please use (the style is already called in the class, ensure that the "woc.bst" style is in your local directory)

% \bibliography{refs.bib} 

\end{document}